# A Study of Energy Band Gap Temperature Relationships for $Cu_2ZnSnS_4$ Thin Films

Prashant K Sarswat,[1,*] Michael L Free[1]

[1]*Department of Metallurgical Engineering, University of Utah, Salt Lake City, Utah, 84112, USA*
[*]*saraswatp@gmail.com*

**Abstract:** The temperature dependent band gap energy of $Cu_2ZnSnS_4$ thin film was studied in the temperature range of 77 to 410 K. Various relevant parameters which explain the temperature variation of the fundamental band gap have been calculated using empirical and semi-empirical models. Amongst the models evaluated, the Varshni and Pässler models show the best agreement with experimental data in the middle temperature range. However, the Bose-Einstein model fits reasonably well over the entire temperature range evaluated. The calculated fitting parameters are in good agreement with the estimated value of the Debye temperature calculated using the Madelung-Einstein approximation and the Hailing method.

---

**References and links**

1. H. Y. Fan, "Temperature Dependence of the Energy Gap in Semiconductors," Phys. Rev. **82** (6), 900-9005(1951).
2. B. Pejova, B. Abay, and I. Bineva, "Temperature Dependence of the Band-Gap Energy and Sub-Band-Gap Absorption Tails in Strongly Quantized ZnSe Nanocrystals Deposited as Thin Films," J. Phys. Chem. C **114**(36), 15280-15289 (2010).
3. N. Nepal, J. Li, M. Nakarmi, J. Y. Lin, and H. X. Jiang, "Temperature and compositional dependence of the energy band gap of AlGaN alloys," Appl. Phys. Lett. **87**, 242104 (2005).
4. C. Rincón, S. M. Wasim, G. Marin, and I. Molina, "Temperature dependence of the optical energy band gap in $CuIn_3Se_5$ and $CuGa_3Se_5$," J. Appl. Phys. **93**, 780 (2003).
5. Z. G. Hu, Y. W. Li, F. Y. Yue, J. H. Chu, and Z. Q. Zhu, "Temperature dependence of optical band gap in ferroelectric $Bi_{3.25}La_{0.75}Ti_3O_{12}$ films determined by ultraviolet transmittance measurements," Appl. Phys. Lett. **91**, 221903 (2007).
6. P. K. Sarswat, M. L. Free, and A. Tiwari, "A study of increased resistivity of FTO back contact for CZTS based absorber material grown by electrodeposition-annealing route," Mater. Res. Soc. Symp. Proc. Vol. 1315, Boston, USA, 2010.
7. P. K. Sarswat, M. L. Free, and A. Tiwari, "A Factorial Design of Experiments Approach to Synthesize CZTS Absorber Material from Aqueous Media," Mater. Res. Soc. Symp. Proc. Vol. 1288, Boston, USA, 2010.
8. P. K. Sarswat and M. L. Free, "A Demonstration of Sol - gel synthesized Bifacial CZTS-Photoelectrochemical cell," Phys. Status Solidi A, DOI 10.1002/pssa.201127216 (2011).
9. P. K. Sarswat, M. L. Free, and A. Tiwari, "Temperature dependent Raman Study of CZTS ($Cu_2ZnSnS_4$) thin film," Phys. Status Solidi B, DOI 10.1002/pssb.201046477 (2011).
10. P. K. Sarswat, M. L. Free, A. Tiwari, and M. Snure, "CZTS Thin Films on Transparent Conducting Electrodes by Electrochemical Technique," Thin Solid Films (to be published).
11. H. Araki, Y. Kubo, A. Mikaduki, K. Jimbo, W. S. Maw, H. Katagiri, M. Yamazaki, K. Oishi, and A. Takeuchi, "Preparation of $Cu_2ZnSnS_4$ thin films by sulfurizing electroplated precursors," Sol. Energy Mater. Sol. Cells **93**, 996-999 (2009).
12. S. Chen, X. G. Gong, A. Walsh, and S. H. Wei, "Crystal and electronic band structure of $Cu_2ZnSnX_4$ (X=S and Se) photovoltaic absorbers: First-principles insights," Appl. Phys. Lett. **94**, 41903 (2009).
13. P. A. Fernandese, P. M. P. Salomé, and A. F. D. Cunha, "Growth and Raman scattering characterization of $Cu_2ZnSnS_4$ thin films," Thin Solid Films **517**, 2519-2523 (2009).
14. P. A. Fernandese, P. M. P. Salomé, and A. F. D. Cunha, "$Cu_xSnS_{x+1}$ (x = 2, 3) thin films grown by sulfurization of metallic precursors deposited by dc magnetron sputtering," Phys. status solidi C **7**, 901-904 (2010).

15. N. Kamoun, H. Bouzouita, and B. Rezig, "Fabrication and characterization of Cu2ZnSnS4 thin films deposited by spray pyrolysis technique", Thin Solid Films **515**, 5949-5952 (2007).
16. Y. P. Varshni, "Temperature dependence of the energy gap in semiconductors," Physica (Utrecht) **34**, 149-154 (1967).
17. R. Pässler, "Basic Model Relations for Temperature Dependencies of Fundamental Energy Gaps in Semiconductors," Phys. status solidi B **200**, 155-172 (1997).
18. P. Lautenschlager, M. Garriga, S. Logothetidis and M. Cardona, "Interband critical points of GaAs and their temperature dependence," Phys. Rev. B **35**, 9174-9189 (1987).
19. A. Manoogian and A. Leclerc, "Determination of the dilation and vibrational contributions to the indirect energy band gap of diamond semiconductor," Can. J. Phys. **57**, 1766-1769 (1979).
20. A. Manoogian and J. C. Wooley, "Temperature dependence of the energy gap in semiconductors," Can J. Phys **62**, 285-287 (1984).
21. G. H. Moh, "Phase relations and mineral assemblages in the Cu-Fe-Zn-Sn-S system," Chem. Erde **34**, 1-61(1975).
22. S. Fiechter, M. Martinez, G. Schmidt, W. Henrion, and Y. Tomm, "Phase relations and optical properties of semiconducting ternary sulfides in the system Cu–Sn–S," J. Phys. Chem. Solids **64**, 1859-1862 (2003).
23. E. Madelung, "Molekulare Eigenschwingungen," Phys. Z. **11**, 898 (1910).
24. A. Einstein, "Comment on Eötvös's law," Ann. Phys. (Leipzig) **34**, 170 (1911).
25. J. Cáceres and C Rincón, "Debye Temperature of $A^IB^{III}C^{VI}_2$ Chalcopyrites and $CuB^{III}_3C^{VI}_5$ and $CuB^{III}_5C^{VI}_8$ Ordered Defect Compounds," Phys. status solidi B **234** (2), 541-552 (2002).
26. T. Hailing, G. A. Saunders, W. A. Lambson, and R. S. Feigelson, "Elastic Behavior of the Chalcopyrite $CdGeAs_2$," J. Phys. C: Solid State Phys. **15**, 1399-1418 (1982)
27. M. V. Yakusheva, A. V. Mudryi, I. V. Victorov, J. Krustok, and E. Mellikov, "Energy of excitons in $CuInS_2$ single crystals," Appl. Phys. Lett. **88**, 011922 (2006).
28. J. Yang, Y. Q. Gao, J. Wu, Z. M. Huang, X. J. Meng, M. R. Shen, J. L. Sun, and J. H. Chu, "Temperature dependent optical properties of Mn doped (Pb,Sr)TiO3 ferroelectric films in absorption region: Electron–phonon interaction," J. Appl. Phys. **108** (111), 114102 (2010).

## 1. Introduction

Temperature dependence of the electronic interband transition can provide valuable information concerning excitonic effects and electron-phonon interactions. Novel research, which discusses effect of lattice vibration and the resulting shift of the energy levels for silicon, was reported by H. Y. Fan, decades ago [1]. Since then, significant research has been accomplished to evaluate the effect of temperature on various semiconductors and photovoltaic materials. For example, optical absorption dependence on temperature for strongly quantized, low dimension semiconductor was reported by Pejova *et al* [2]. Nepal *et al* [3] utilized deep ultraviolet photoluminescence spectroscopy to study band gap temperature dependence in AlGaN. Rincon *et al* [4] reported energy band gap temperature dependence of bulk crystals of $CuIn_3Se_5$ and $CuGa_3Se_5$. Hu *et al* reported band gap measurements for films grown by chemical solution based methods [5]. Recently, $Cu_2ZnSnS_4$ (CZTS) and similar quaternary chalcogenides have received considerable attention due to a high absorption coefficient ($\sim 10^4$ cm$^{-1}$) and an optimal fundamental band gap energy ($\sim$ 1.5 eV) [6-15]. However, the authors are not aware of research reported regarding the band gap temperature dependence of CZTS absorber material. It is important to note that the near band edge transition with respect to temperature is an important parameter for optimizing the performance with respect to operating temperatures of CZTS based photovoltaic devices, which can vary significantly daily and seasonally.

Amongst reported band gap models three popular models have been evaluated in this study: (1) the Varshni model [16], (2) Pässler model [17] and (3) the Bose Einstein model [18]. These models suggest that band gap shift in semiconductors as a function of temperature is mainly due to electron – phonon interactions. However, thermal expansion also plays a minor role, which was considered by several research groups. For example, Manoogian *et al* [19-20] used a modified Varshni model to consider the effects of lattice expansion, and Rincon *et al* [4] used the contribution of lattice dilation to evaluate the shift of band gap energy. In view of these issues, a study of fundamental energy band gap variation of CZTS thin films, (grown by solution based technique) as a function of temperature has been performed. UV-Vis spectroscopy was utilized to measure band gap parameters. A comparative study between different models has been performed in order to determine the best model for CZTS from 77 to 400 K.

## 2. Experimental

CZTS thin films were grown on a conducting transparent substrate by a solution based method. Cu-Zn-Sn layers were electrochemically grown. The Cu layer was deposited using a $CN^-$ medium bath whereas zinc and tin were deposited using a $Cl^-$ medium bath [10]. Thin films, coated with metal precursors (Cu, Zn and Sn), were sulfurized by annealing in a sulfur environment. Annealing was performed in an argon environment with evaporated elemental sulfur (99.99%) in a tube furnace for two hours at ~ 560$^o$ C. Fig. 1 shows a characteristic θ-2θ XRD pattern for a film after sulfurization. This pattern matches well with the kesterite structure of CZTS (JCPDS card 26-0575). Peaks corresponding to (112), (103), (200), (105), (220), (312), (224), (314), (008) and (332) planes were detected, revealing the polycrystalline nature of CZTS film. Raman analysis was also performed to distinguish phases. All three major peaks, which correspond to the CZTS phase, can be seen in the inset of Fig. 1 [8]. No distinct peaks corresponding to impurity copper sulfide, zinc sulfide, and tin sulfide were observed. The lack of peaks or shoulders for copper tin sulfide (CTS) at 355, 348, 351, and 295 cm$^{-1}$ provides additional evidence of CZTS film purity [14].

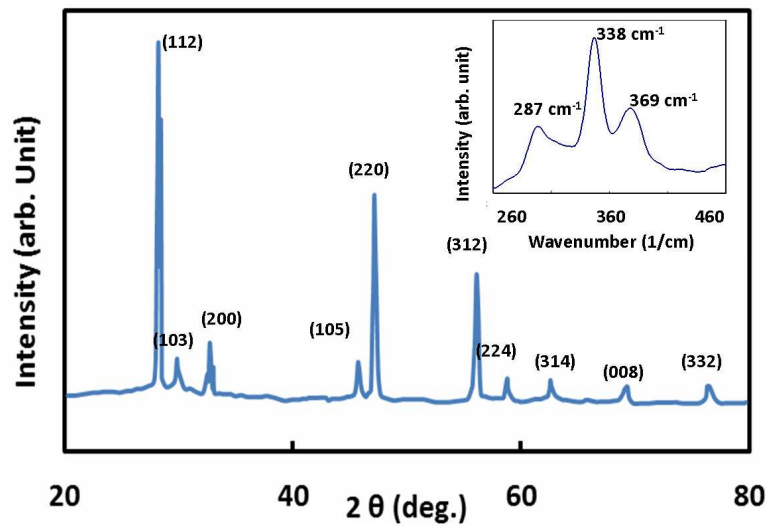

**Figure 1.** XRD pattern of CZTS thin film used for temperature dependent energy band gap measurements. All peaks closely match with kesterite crystal structure (JCPDS no. 26-0575); Raman spectrum of CZTS thin film (shown in inset), three distinct peaks, corresponding to kesterite crystal structure can be seen.

Prior to transmittance measurements, elemental and morphological characterization of film was carried out using various techniques such as scanning electron microscopy, EDS, atomic force microscopy and ICP spectroscopy (Supplement Figures included). All elemental characterization reveals that the elemental ratio of Cu:Zn:Sn is close to 2:1:1. Melting point testing showed that the film melted far above the melting point of CTS (~ 735$^o$C) and very near the melting point of CZTS (~ 820$^o$C) [21, 22]. These observations suggest that films are relatively pure for band gap measurements. The surface morphology of film is found to be smooth with an average roughness of ~ 30 nm in most regions of film. Temperature sensitive transmittance and reflectance measurements of CZTS film were done using a cryostat setup coupled with an Ocean optics spectrophotometer, which is capable of measurements from 77 K to 500 K. Transmittance and reflectance data were utilized for band gap energy evaluation [15]. The band gap energies of CZTS films were estimated by extrapolating the linear portion of graphs of $(\alpha E)^2$ versus energy to zero (Fig. 2). The band gap at 0 K was obtained by extrapolating the band gap obtained near 77 K, using various models. Optical absorption spectra (and Tauc plot) generally reveal some evidence of defects (such as disordered semiconductor, joint density of states, indirect semiconductor property) by the presence of a largely exponential Urbach tail when defects are present. The authors carefully measured all transmittance data and did not find evidence of such defects. The observation also suggests that likelihood of weak bond density is also low. Possible exponential dependence of alpha on the photon energy was examined. Results show that possibility of the disorder is low.

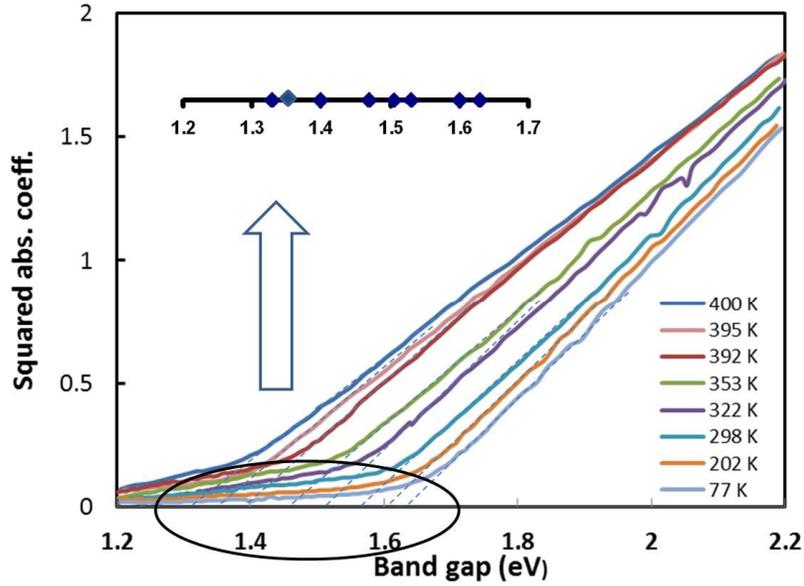

**Figure 2.** Squared absorption coefficient vs. band gap energy is shown for CZTS thin film at different temperatures.

## 3. Results and Discussions

The effect of temperature on band gap energy shrinkage has been quantified through several empirical or semiempirical relations. Among the empirical relationships, the Varshni relationship is often used to assess nonlinear temperature dependent band gap shift. The Varshni equation [16] for the band gap variation as a function of temperature is:

$$E_g(T) = E_g(0) - \frac{AT^2}{T+B} \qquad (1)$$

Where A and B are fitting parameters, which are characteristic of a given semiconductor. $E_g(0)$ is the band gap of the semiconductor at 0 K. The Varshni relation is a combination of quadratic low temperature asymptotic behavior with linear high temperature dependence [2, 3]. Although this relation fits appropriately for various III-V and II-VI semiconductors, it gives negative values for the fitting parameters (A and B) for various known wide band gap semiconductors. Moreover, the physical significance of parameter B was not explained, which is believed to be related with the Debye temperature of the material [2, 3]. Fig. 3 shows experimental band gap shift as well as fitting using equation (1), based on the Varshni model. The values of fitting parameters A and B are $10.10 \times 10^{-4}$ eVK$^{-1}$ and 340 K respectively, whereas $E_g(0)$ was 1.64 eV. It can be seen that the Varshni relation fits well for low temperatures (up to 360 K); however, it shows deviation from experimental values, above 360 K.

Another model, proposed by Pässler, considers a power law type spectral function which is related to electron – phonon interaction [17]. This model explains curvilinear behavior of band gap shift at moderate temperatures. It accommodates band gap shift resulting from contributions of phonons with various energies. The expression for band gap shift is given by [17]:

$$E_g(T) = E_g(0) - \frac{\alpha \Theta}{2}\left[\left(1+\left(\frac{2T}{\Theta}\right)^q\right)^{1/q} - 1\right] + \text{Contribution of thermal dilation} \quad (2)$$

Here $\Theta$ is a constant closely related with average phonon temperature, the parameter $q = \eta + 1$, where $\eta$ is shape factor of spectral function, and $\alpha$ is the limiting slope of band gap shift at high temperatures [16]. Fig. 3 includes experimental band gap and Pässler model data. The contribution of thermal dilation is ignored for simplicity. This model also fits well below 355 K; however significant deviation can be seen at temperatures above 350 K. The values of adjustable parameters obtained are: $\Theta$ = 260 K; q = 2.7; $\alpha$ = 7.7 × 10$^{-4}$ eVK$^{-1}$. The Debye temperature, which is proportional to $\Theta$, can also be estimated by a method described by Rincon *et al.*[4] The estimated Debye temperature is ~ 360 K.

The Bose-Einstein model, which considers electron interaction within crystals, also relates energy shift and temperature with Debye energy [18]. According to this model, the band gap energy can be determined from:

$$E_g(T) = E_g(0) - \frac{2a_B}{\exp\left(\frac{\Theta_E}{T}\right) - 1} \quad (3)$$

Where $a_B$ is a parameter associated with the strength of exciton – phonon interactions within the crystal; $\Theta_E$ is average temperature (Einstein characteristic temperature) of phonons interacting with the electronic subsystem. The Debye temperature for an Einstein oscillator can be determined from the relationship: $\Theta_D = 4/3\,\Theta_E$ [2]. Although the Debye temperature or Einstein temperature for CZTS based compounds or other similar quaternary chalcogenides is not readily available, available data for ternary compounds can be utilized for a rough estimation. According to Madelung [23] and Einstein [24], the Debye temperature $\Theta_D$ for solids can be expressed in term of bulk modulus:

$$\Theta_D = C_B \left(\frac{h}{k_B}\right)\left(\frac{B_o V^{1/3}}{M}\right)^{1/2} \quad (4)$$

Where $C_B$ is a dimensionless parameter, $K_B$ and $h$ are Boltzmann's and Plank's constants respectively, $V$ is mean atomic volume of lattice, $M$ is mean atomic weight per site and $B_0$ is bulk modulus. Cáceres *et al* [24] used this equation for calculation of the Debye temperature. They used a $B_0$ value at $V = 0.024$ nm$^3$ which was found to decrease with increase in $V$. For CZTS, the value of available lattice parameters can be utilized for evaluation of $V$ and $M$ [12].The resulting value of $V$ is utilized to estimate the bulk modulus. The final estimated value of Debye temperature for CZTS is ~ 333 K. The $C_B$ value used for this calculation is ~ $2.85 \times 10^7$ [25].

A similar estimation can also be made based on work of Hailing *et al* [23]. This research explored various elastic constants for chalcopyrite type structures as well as the Debye temperature.

Debye temperature can be calculated by integrating over velocity space using the equation [26]:

$$\Theta_D = \left(\frac{9N}{4\pi V}\right)^{1/3} \left(\frac{h}{k}\right) \left[\int \left(\frac{1}{V_j^3} + \frac{1}{V_k^3} + \frac{1}{V_l^3}\right) \frac{d\Omega}{\pi}\right]^{-1/3} \quad (5)$$

Where $V_j$, $V_k$, and $V_l$ are eigenvalues of the Christoffel equation, $N/V$ corresponds to number of atoms per unit volume of lattice, $\Omega$ is the solid angle, $k$ is the Boltzmann constant, and $h$ is Plank's constant [26]. The value of the integral inside the bracket was calculated for known material such as CdGeAs$_2$ and is utilized to estimate the Debye temperature for CZTS. The $N/V$ value for CZTS can be calculated based on reported data [12]. Based on equation (5), the Debye temperature for CZTS is ~ 349 K, which is close to that obtained from the Madelung-Einstein approximation (~ 333 K). Fig. 3 shows the theoretical band gap shift using equation (3) together with the experimental band gap shift. It can be seen that this model shows the best fit for the entire temperature range examined. In contrast, other models reveal a significant deviation from measured data above 350 K. The value of fitting parameter a$_B$ obtained from least squares fitting of the model to the measured data is 90.25 meV.

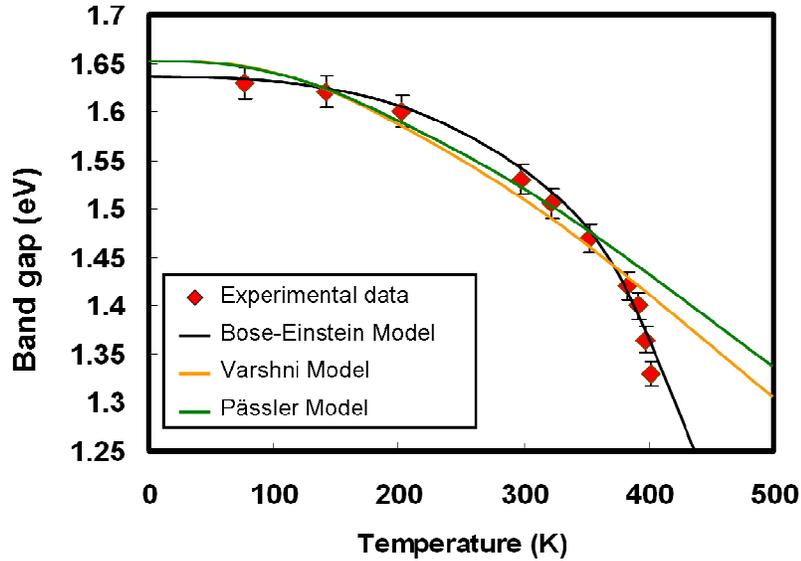

**Figure 3.** Experimental band gap energy vs. temperature; Fitting based on Varshni model, Pässler model, and Bose-Einstein model is shown by continuous lines of different colors. A very good fitting can clearly be seen over the entire temperature range in the case of the Bose-Einstein model.

The Varshni equation is a second order approximation to the Bose –Einstein model and valid for certain range of temperature [19, 20]. However, Bose –Einstein expression, a more generalized expression, best describes band gap shrinkage behavior of $CuInS_2$, whose crystal structure is quite similar to CZTS [27]. Bose –Einstein best fit for CZTS absorber suggests that small shift of atoms or ion from balance sites (arises due to lattice vibration) will diminish lattice periodic field which further affect chemical bond length and band gap energy [28]. The results also suggest that increase in temperature enhances activated population of phonon in CZTS absorber material, which in turn causes more electron-phonon interaction. This interaction results in band gap shrinkage at elevated temperatures. Another possibility is domination of the dynamic part of electron-phonon coupling up to a certain range of temperature, which is not considered in the extended range of temperature in the Varshni equation. The Pässler model which considers some contribution of thermal dilation was ignored in fitting. Incorporation of this contribution will certainly improve fitting. The average band gap narrowing coefficient ($dE_g/dT$) was ~ - $8.63\times10^{-4}$ eV/K (Bose-Einstein Model), -$5.8 \times 10^{-4}$ eV/K (Pässler model), and -$6.6\times10^{-4}$ eV/K (Varshni Model) respectively. A summary of fitting parameter is listed in Table 1.

Table 1. List of various fitting parameters

| Serial No. | Model or Equation | Fitting Parameter | Fitting Parameter | Debye Temperature | ($dE_g/dT$) |
|---|---|---|---|---|---|
| 1 | Varshni Model | A=$10.10 \times 10^{-4}$ eVK$^{-1}$ | B=340 K | | -$6.6\times10^{-4}$ eVK$^{-1}$ |
| 2 | Pässler model | q = 2.7 | $\alpha$ = $7.7 \times 10^{-4}$ eVK$^{-1}$ | 360 K | -$5.8 \times 10^{-4}$ eVK$^{-1}$ |
| 3 | Bose - Einstein model | $a_B$ = 90.25 meV | | | -$8.63\times10^{-4}$ eVK$^{-1}$ |
| 4 | Equation 4 | | | 333 K | |
| 5 | Equation 5 | | | 349 K | |

## 4. Concluding Remarks

In summary, the fundamental band gap relationship with temperature for CZTS thin film was measured using transmittance data from 77 to 410 K. Band gap data shows inadequate fitting with the Varshni model and Pässler model. The Bose - Einstein model best describes the band gap temperature dependence for CZTS, which can be understood as existence of carrier-phonon coupling in CZTS. Estimated Debye temperatures obtained from various theories and approximations lie in close proximity to Debye temperatures obtained from fitting parameters determined from best fit Bose - Einstein and Pässler models. This information provides useful data for predicting the effect of temperature on CZTS band gap for solar cell applications that are affected by daily temperature variations.

**Acknowledgments**

The authors want to thank Jiajia for providing liquid Nitrogen for low temperature measurements.